\documentclass[conference,a4paper]{IEEEtran}
\IEEEoverridecommandlockouts 
\ifCLASSINFOpdf
\usepackage[pdftex]{graphicx}
\graphicspath{{./Images/}}
\DeclareGraphicsExtensions{.pdf,.jpeg,.png}
\else
\fi

\usepackage[cmex10]{amsmath}
\usepackage{siunitx}
\usepackage{cite}
\usepackage{psfrag}
\usepackage[utf8]{inputenc}
\usepackage[T1]{fontenc}
\usepackage{amsmath,amssymb,amsthm,mathrsfs,amsfonts,dsfont} 
\usepackage{mathabx}
\usepackage{mathrsfs}
\usepackage[nolist]{acronym}
\usepackage{amssymb}
\usepackage{amsmath}
\usepackage{graphicx}
\usepackage{cite}
\usepackage{multirow}
\usepackage{wasysym}
\usepackage{float}
\usepackage{color}
\usepackage[utf8]{inputenc}
\usepackage{lmodern,textcomp}  
\usepackage{tabularx} 
\usepackage[hyphens]{url}
\usepackage{graphicx} 

\title{Agent-based Model for Spot and\\ Balancing Electricity Markets}

\author{
\IEEEauthorblockN{Florian Kühnlenz, Pedro H. J. Nardelli}
\IEEEauthorblockA{Centre for Wireless Communications (CWC)\\ 
	University of Oulu, Finland\\
Contact: [Florian.Kuhnlenz,Pedro.Nardelli]@oulu.fi}

\thanks{This work is partly funded by Finnish Academy (n. 271150) and CNPq/Brazil (n.490235/2012-3) as part of the joint project SUSTAIN, by Strategic Research Council/Aka BC-DC project (n.292854) and by the European Commission through the P2P-SmarTest project (n.646469).}%
}

\begin{document}

\maketitle

\begin{abstract}

We present a simple, yet realistic, agent-based model of an electricity market.
The proposed model combines the spot and balancing markets with a resolution of one minute, which
enables a more accurate depiction of the physical properties of the power grid. 
As a test, we compare the results obtained from our simulation to data from Nord Pool.

\end{abstract}

\section{Introduction}

In the electricity grid, it is necessary that supply and demand are in balance all the time, due to physical constraints \cite{nardelli2014models}.
The main part of power generation still happens with synchronous generators.
They can only slightly deviate from the nominal frequency for short amounts of time without damage. To ensure that supply and demand match, electricity is traded in several stages with increasing time resolution. The two most important trading stages are the daily spot market and the balancing market \cite{nordpool:blc}. All differences that could not be accounted for in the spot market need to be corrected at the balancing market in real time. 

Due to the still coarse time resolution of most spot markets (usually one hour) they cannot accurately predict how suppliers need to run their power plants to match demand all the time. Therefore, a simulation with high time resolution is needed to accurately model the behavior of balancing markets. 

Nevertheless, even though balancing markets have become more and more an issue of interest (e.g. \cite{Fingrid:2016us,Hirth:2013td,Ocker:2015wg}), there is almost no tools available to get insights in the interplay between balancing and spot markets. Yet, as demonstrated in \cite{Weiss:2009,Li:2011} the design of the spot market has great impact in the balancing market the physical behavior in the grid.
Most openly available simulation cover power flow and unit commitment models \cite{PyPSA,Greenhall:2012wz}, the general energy system \cite{EMMAmodel,ficus} and often include at least an approximation of a spot market. Even when balancing markets are discussed they are often analyzed individually \cite{Santos:2012hv,Mureddu:2015ck}.

Furthermore in real markets a lot of parties take part in the trading process already, with the amount most likely rising in the future. Due to generation becomes more decentralized with renewable sources and consumption and production happening in new structural units like micro-grids \cite{nardelli2014models}. 

To better understand the interplay between players, the proposed model is designed as an agent based model (ABM)\cite{Helbing:2011th}, in which multiple agents of each type (e.g. producers or utilities) may co-exist. 
The trading of electricity is an integral part of the whole system and shall become even more important with the ongoing introduction of the smart grid concept. The market price of electricity is, however, often assumed to be an externality in models for smart grid communication or physical grid simulations. 

In contrast, the flow of information, in form of real time prices, is assumed to change the way that electricity is consumed, e.g. in demand response systems. If the consumption is changed by the price, it will inevitably change the price in-turn since it reflects the point where consumption and production meet. To better understanding these interactions, one needs to include these intrinsic market aspects in simulations instead of assuming the price as something external.
Targeting this issue, We introduce here a simple simulation model of the electricity market that takes this internal relation into account.

\section{Liberalized electricity market structure}

The trading of electricity usually takes place in several stages (e.g \cite{nordpool:blc}) as to be described next.

\subsection{Long-term}

In the first one, long term contracts are made between two parties often covering the time span of multiple years. These contracts cover the so-called base-load, the very stable and predictable part of electricity consumption. 

\subsection{Spot Market}

In the next stage, often called sport market, electricity is usually traded in a time span of a day with an hourly time resolution between multiple parties and a market maker. Since these trades usually end several hours before the actual delivery hour, there needs to be at least one more trading stage to cover any changes that happen between the end of trading and the actual delivery. 

Sport markets are typically operated with a single price for each hour. Each party submits its bids for each hour of the day. The bids describe either a bid to buy a certain amount of power or to produce a certain amount of power for a certain price. The market maker is the only one who will see all offers and is therefore able to determine which price will optimize the social welfare of all parties. This price then determines which producer must produce what amount of power during any given hour and consequently how much can be consumed. It is important to remember that the consumption is not controllable; rather it is based on forecasts, as is the production of renewable energy sources like solar and wind. 

Since the demand (and production to a growing extend) that is traded in the spot market stage is based on forecasts, the importance of the market stages with higher time resolution and shorter lead times has become more important over the years\cite{Ocker:2015wg}.

\subsection{Intraday and balancing market}

Usually there are two more stages: one before the delivery hour, often called intraday market, and one final, so called balancing market stage. The balancing takes place during the delivery hour and aims to ensure that supply and demand are matched as close a possible. The balancing market can be split in two timescales again, one where power is delivered typically within 15 minutes and one where power is delivered instantaneous depending on the physical conditions of the grid for up to 15 minutes. This last stage happens distributed among the participating generators and will be compensated for afterwards.

The balancing and intraday markets cannot only be used to compensate for updated forecasts, but can also cover an outage of a power plant. The third job of these market stages is to fine tune the production curve on a finer timescale.  

It is for example possible that the demand for a given hour is as predicted, yet large amounts of balancing power are needed\cite{Weiss:2009}. Since the spot market is based on a hourly time resolution, the participants can only agree on the power needed during the full hour in aggregate. If for example most of the power is needed during the second half of the hour, there will be down-regulation needed in the first part of the hour and up-regulation in the second. If the intraday market provides sufficiently high time resolution this would mean that power would be sold during the first part of the hour and the same amount needed to be bought for the second. This intrahour balancing is a significant cause for balancing\cite{Weiss:2009,Li:2011}.

The working of intraday markets can differ between different market implementations and is not as equal as the spot market design. However, all intraday markets close before the delivery hour in contrast to balancing markets. In balancing markets the bids are submitted ahead of time, however are called upon during the delivery hour when needed. While in the balancing and in the spot market it will be settled before the delivery hour which bids and offers are matched. Therefore, the price of balancing is only known after the delivery hour is over and it is known which up- or down regulation bids were used. Opposite to the spot market or intraday market prices which are known before the power is produced and consumed.

\section{Proposed model}

The presented model of an electricity market aims to provide the basic concepts of a working electricity markets with all major participants in a simple form. This serves two main purposes: 1. it helps to understand the interaction between all parts of the model and second it provides the freedom to specialize the model for specific purposes later. As a side effect, it should also speed up execution time. 
As a drawback of this approach, the basic form the model cannot predict specific outcomes but only general behavior. 
If any details in the implementation needed to be very specific it was modeled as close as possible to the Nord Pool market and therefore to the underlying matching algorithm Euphemia\cite{Euphemia}.

The Euphemia algorithm is in fact used across most European markets, making the model quite universal.
The model is constructed as an agent based model\cite{Helbing:2011th} (ABM) where all market participants are agents. Agent based models are used in a variety of contexts\cite{Macal:2016kh} and in the context of power grids \cite{Kremers:2013vr,Kremers:2010,Shafiei:2012cj,Kuhnlenz:2016iz}. In the context of economic research, more conservative approaches are, however, the mostly employed, which has been recently criticized \cite{Farmer:2009cd}.

A big advantage of agent based models in this context is that they are not confined to equilibrium states and more accurately capture the complexity of economic realities\cite{van2012agent,Cristelli:2015ga}. 
More specifically agents make individual decisions based on their perceived environment and their internal state. The state of an agent might change during the simulation based on certain rules so that the decisions of an agent influence the environment, which in-turn influences the state of the agent which again influences the decisions. This makes it possible to capture basic feedback loops and dynamic behavior.	

In our proposed model, we employ the following types of agents:

\begin{itemize}

\item \textbf{Producers:} provide power with for a given price per MWh. Producers have a maximum capacity of how much power they can deliver in every time step. Every producer can bid a certain amount of his power into the balancing market.
\item \textbf{Utilities:} a utility forecasts and buys the power for its assigned users for the next day and distributes balancing costs among the users.
\item \textbf{Users:} have a certain power demand during the day which might change due to prices or other internal or external factors.
\end{itemize}

\subsection{Producers}
The basic producer’s agents are simple. Their price and capacity is fixed. The provide offers for the market and keep track of their production schedule according to the market. They also offer a certain amount for balancing based on globally set percentage or an internal on.
There are currently also two sub-classes of producers for simulating wind and solar production. Both make an internal forecast about their production that is offered on the market. The realized output of both is not exactly according to the market set schedule but can differ up or down, therefore requiring balancing. Solar production has its peak always at the same time, while wind production peaks can occur randomly during the day.
Producers keep track of their income from spot and balancing markets, and balancing payments when they are producers of renewables.

\subsection{Utilities}
The utility agent forecasts the consumption of its assigned users based on their prior usage. Therefore, the utility keeps track of up to 30 days of aggregated usage data and calculates a weighted average of the data to forecast the next day. The forecast is also multiplied with an error. The error is modeled as a random walk with mean return. 
The utility keeps track of revenue from its users and costs for buying at the spot and balancing markets. Since the balancing costs cannot be attributed to specific users they are shared by all users. Therefore, the fix costs of the utility are calculated as:
\[(revenue-cost-balancing\_cost)/number\_of\_users\]

\subsection{Users}
\begin{figure}[t]
      \centering
      \includegraphics[width=\columnwidth]{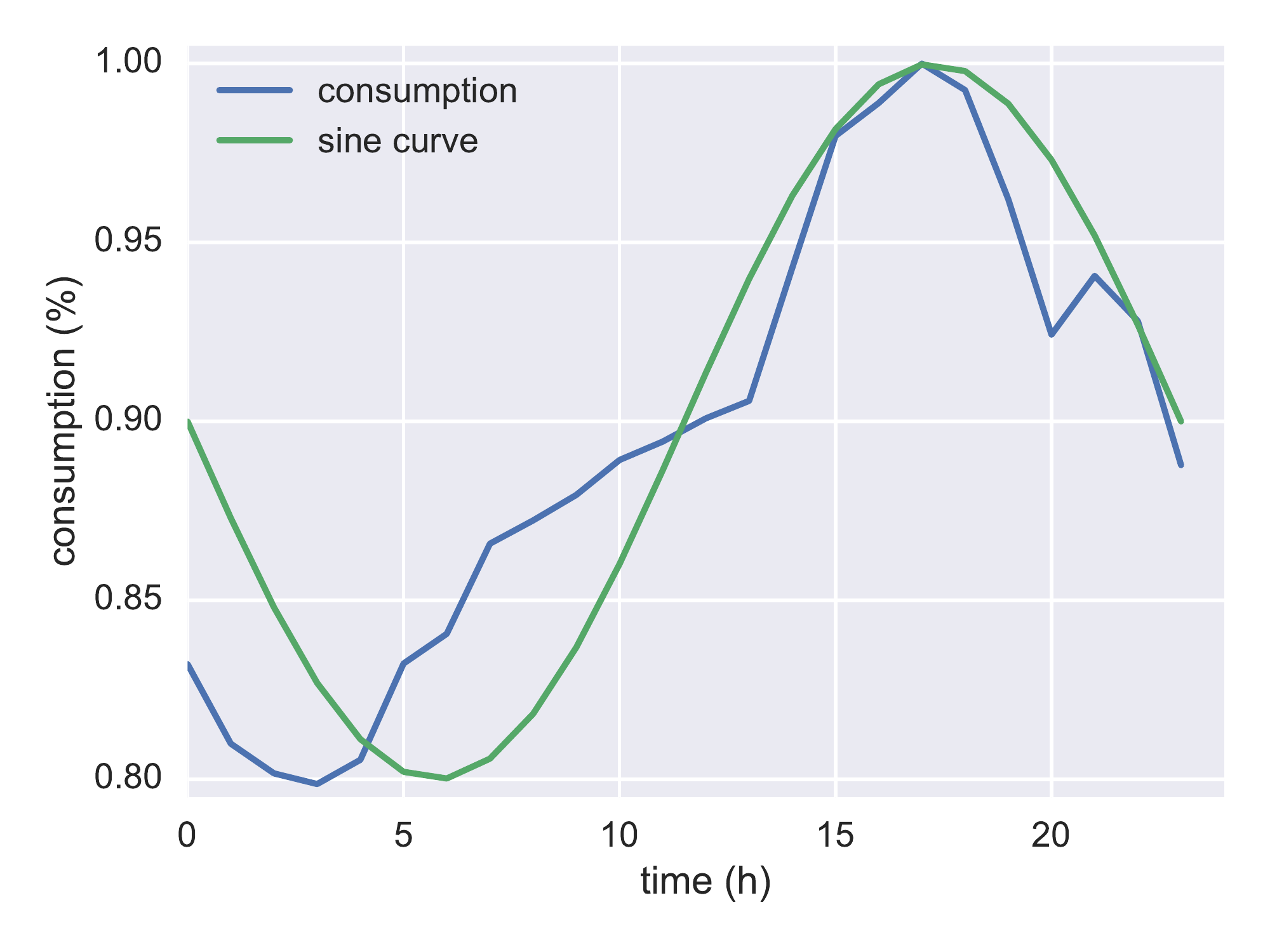}
      \caption{Comparison of Finnish consumption during a winter day with a sine curve.}
      \label{sine}
   \end{figure}

The user agent mainly generates a load curve currently based on the ``\(\sin\)'' function. It has a fixed minimum and maximum value however the phase of the sine curve changes during the simulation. The maximum of the sine curve is occurring at around 6 p.m. in the evening. A comparison of a sine curve with the actual consumption pattern for a winter day can be seen in Fig. \ref{sine}.
There are two types of users, optimizing users and normal users. For normal users, the curve might shift randomly by up to 15 minutes in any direction. The optimizing user can shift his sine curve freely. This allows him to use the price as an input to optimize his daily usage.

\subsection{Simulation Periods}
\begin{figure}[t]
      \centering
      \includegraphics[width=\columnwidth]{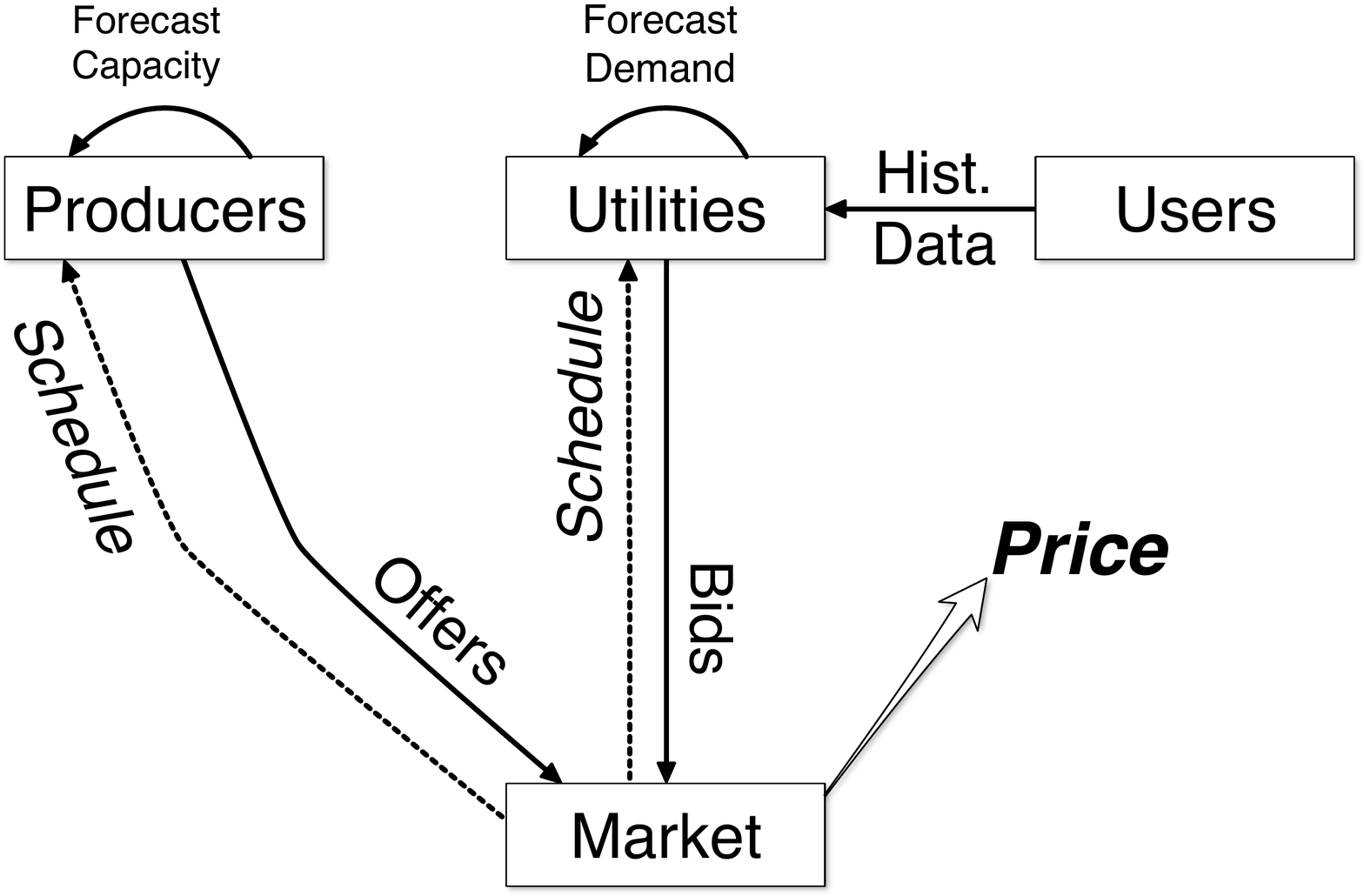}
      \caption{Spot market period. First time step: Utilities forecast the demand of their users and submit bids accordingly; producers forecast production capacity and submit offers accordingly. Second time step: market matches bids and offers to create schedules and a public price.}
      \label{spotPeriod}
   \end{figure}

\begin{figure}[t]
      \centering
      \includegraphics[width=\columnwidth]{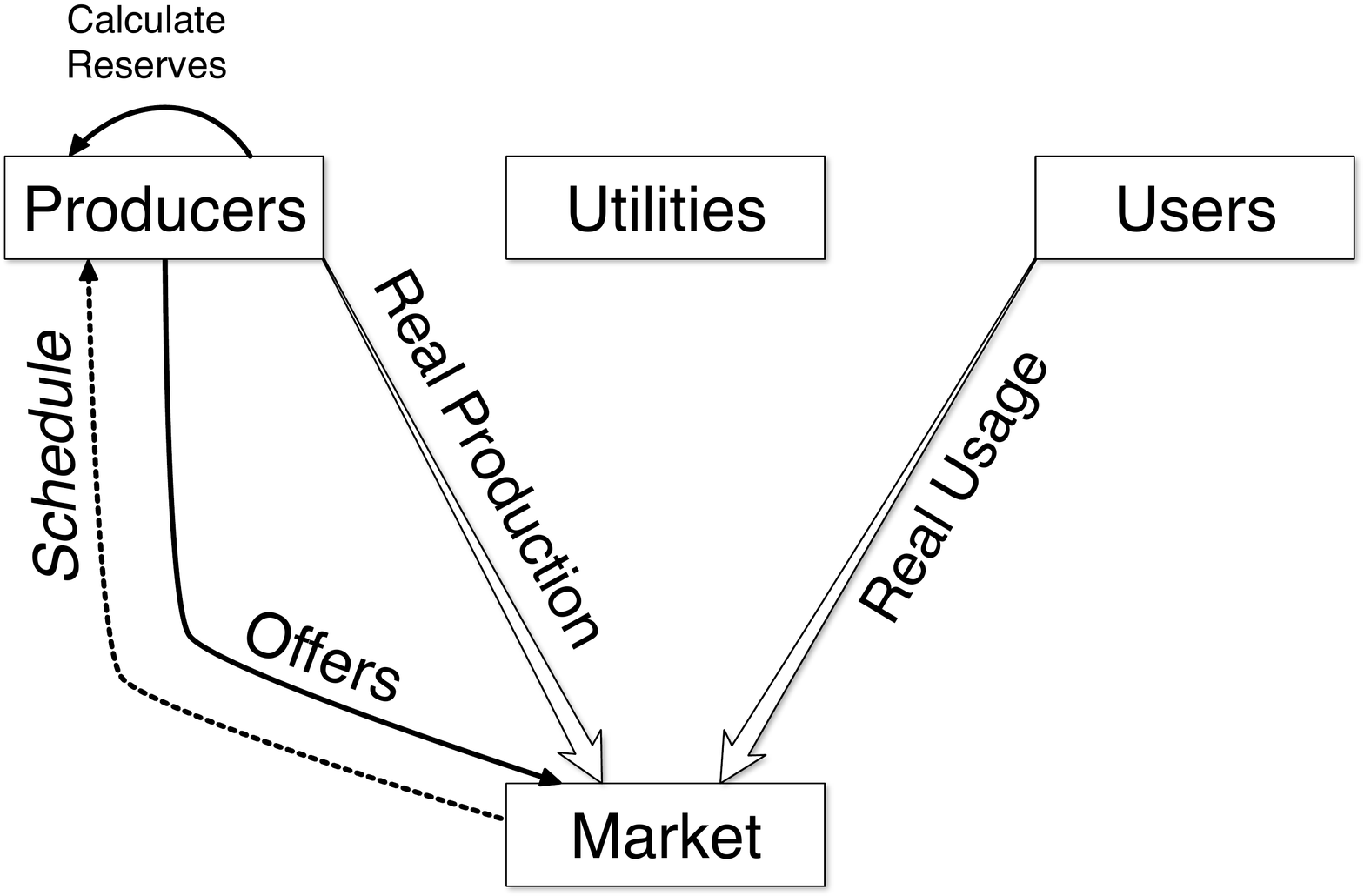}
      \caption{Balancing period. Producers submit balancing offers, which the market calls upon depending on the mismatch between real usage and production.}
      \label{balancingPeriod}
   \end{figure}

\begin{figure}[t]
      \centering
      \includegraphics[width=\columnwidth]{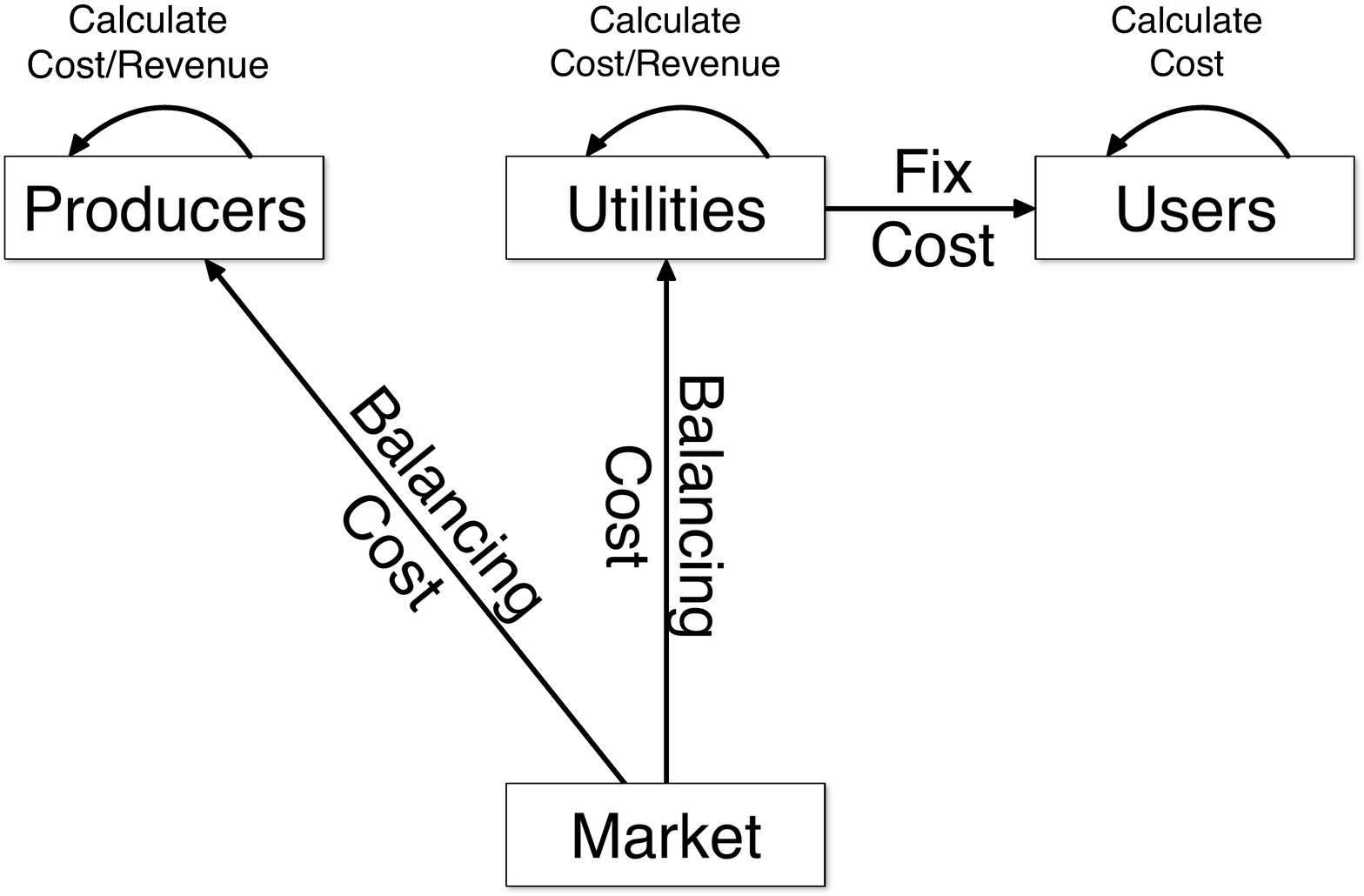}
      \caption{After market period. All players update their costs and revenues. The market collects balancing payments and pays for called offers. Balancing costs of utilities are passed to the user as fixed costs.}
      \label{afterMarket}
   \end{figure}

Every simulation day occurs in three main stages depicted in Figs. \ref{spotPeriod}, \ref{balancingPeriod} and \ref{afterMarket}. During the spot market period, all producers submit their offers and utilities their bids according to their forecasts. In the current implementation utilities have no price flexibility so their bids always need to be matched, which in the real market would be represented by the bid having the maximal allowed price. 
During the balancing period, all balancing offers are collected and then called upon if needed, starting with the cheapest ones.

The differences between production and consumption are then compared for every given cycle, typically 15 minutes. If the difference exceeds a certain amount of power, an up- or down- regulation offer is called upon. The power plant is then removed from the list of both up and down regulation, to minimize the fast changes in output, which are typically  though on the equipment. 

The adjustments are valid for the full rest of the hour.
For example, if a power plant must adjust its output down during the first 15 minutes of an hour, it will remain at that output till the next full hour starts.
In the final period, the balancing costs for every period are calculated according to the pricing scheme that is used in Nord Pool\cite{fingrid:blc,nordpool:blc}. Additionally, all agents account for their consumed or produced power.

\section{Comparison}

To verify the results of the simulation, we provide a comparison with data obtained from the Nord Pool market\cite{nordpoolData}. It is important to state that it is not the goal of the simulation to model a specific market, like Nord Pool in great detail, but rather to capture the general behavior. 

Hence, the simulation was not calibrated with Nord Pool data but rather setup to simulate a much smaller system with comparable prices. However, the results show very comparable outcomes at the balancing market, specially concerning the intrahour balancing that could only be captured due to the high time resolution of the simulation.

In Table \ref{comparison} the comparison between the data from Nord Pool and a 30-day run can be seen. Some of the bigger differences might be due to the much shorter run time of the simulation of only 30 days instead of a whole year. However, the simulation does not undergo any seasonal changes therefore the results are expected to be representative. 
In the presented simulations, there was a total of 100.000 users with none of them being optimizing users. All users where spread among 6 utilities, while production was provided by 11 producers.

\begin{table}[b]
\caption{Comparison between Nord Pool and the proposed model}
\label{comparison}
\begin{center}
\begin{tabularx}{\columnwidth}{|X|c|c|}
\hline
 & Nord Pool (2015) &  Simulation (30 days)\\
\hline
avg. price & 21.00€ & 22.86€\\
\(\sqrt[]{\sigma^2}\) & 7.92€ & 10.62€\\
avg. regulation & 1.59\% & 1.10\%\\
max. regulation & 7.14\% & 5.02\%\\
intra-hour regul. & 13.09h & 11.68h\\
balancing price & 171\% / 60\% & 186\% / 64\% \\
\hline
\end{tabularx}
\end{center}
\end{table}

\begin{figure}[t]
      \centering
      \includegraphics[width=\columnwidth]{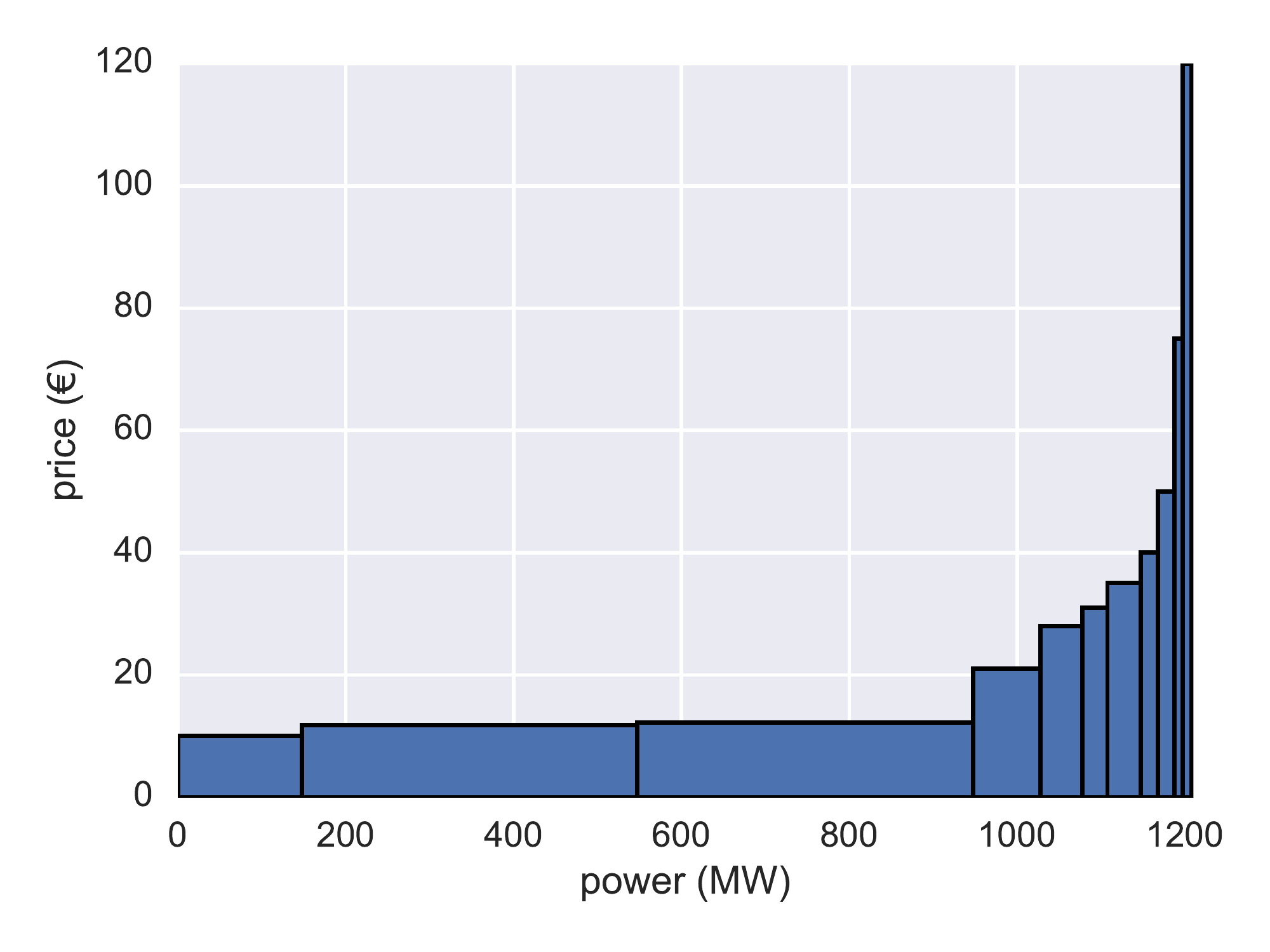}
      \caption{The price and production capacity of the simulated power plants.}
      \label{priceVSpower}
   \end{figure}

Fig. \ref{priceVSpower} shows the power plant configuration used in the simulated system. All bigger power plants have a very low regulation factor and only provide small amounts of power for balancing. This captures the situation that most thermal power plants, which provide base load, are not very flexible. 

\begin{figure}[t]
      \centering
      \includegraphics[width=\columnwidth]{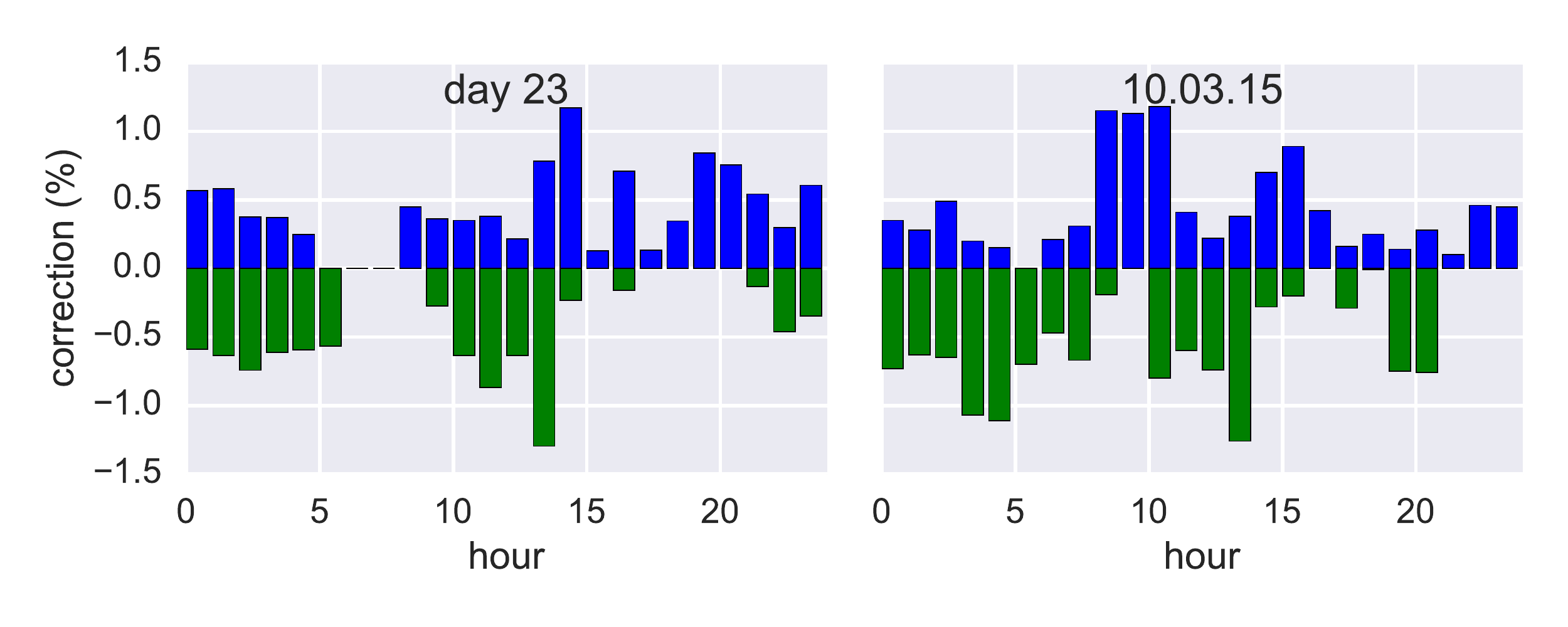}
      \caption{Left: balancing needed for a specific day in the simulation. Right: comparable day from Nord Pool data.}
      \label{balancing}
   \end{figure}

In Fig. \ref{balancing} we see the amount of balancing in relation to the consumption for one day of the simulation compared to a chosen day from the Nord Pool data. The days are specifically picked to be very comparable and highlight the phenomenon of intrahour regulation. For the Nord Pool plot, both automatic and market based regulation where considered, as there is no difference between these in the simulation.

Overall the simulation provides results close to data from Nord Pool, considering the simplifications e.g. no intraday market and only sinusoidal load curves. This shows the model does not only provide all basic functionality but also produces data which is comparable to real markets.  

\section{Conclusions and Outlook}

The presented model of a basic electricity market including spot and balancing markets.
This provides the basis for future research on the interaction between different markets, players and possible other systems like the physical grid or communication networks. 
The high time resolution of the model opens further possibilities for interaction with physical phenomena in the grid and the optimization of market operation times. 

For example, the effect on the balancing or the integration of renewables of shortening the spot market interval to 30 or 15 minutes can be readily tested.
We derived the model from real markets and close to the implementation of Nord Pool and could show that the output is consequently comparable. Yet, the model remains very flexible so it can simulate systems that are spanning several countries or just a micro-grid.



\end{document}